\documentclass[usenatbib]{mn2e}
\usepackage{epsfig}
\usepackage{amsmath}
\usepackage{ulem}

\def\gsim{\mathrel{\raise0.35ex\hbox{$\scriptstyle >$}\kern-0.6em 
\lower0.40ex\hbox{{$\scriptstyle \sim$}}}}
\def\lsim{\mathrel{\raise0.35ex\hbox{$\scriptstyle <$}\kern-0.6em 
\lower0.40ex\hbox{{$\scriptstyle \sim$}}}}
\def\gs{\mathrel{\raise0.35ex\hbox{$\scriptstyle >$}\kern-0.45em 
\lower0.40ex\hbox{{$\scriptstyle \sim$}}}}
\def\ls{\mathrel{\raise0.35ex\hbox{$\scriptstyle <$}\kern-0.45em 
\lower0.40ex\hbox{{$\scriptstyle \sim$}}}}
\def\lya{{\rm Ly$\alpha$}}
\def\ha{{\rm H$\alpha$}}

\def\msun{{\rm M}$_{\odot}$}

\date{Accepted 2007 March 14. Received 2007 March 9; in original form 2007 January 8}

\title[First Appearance of Red Sequence in Proto-Clusters]
{The First Appearance of the Red Sequence of Galaxies in Proto-Clusters at $2\ls z\ls3$}

\author[T. Kodama et al.]{
\parbox[t]{\textwidth}{
Tadayuki Kodama$^{1}$\thanks{E-mail: kodama@optik.mtk.nao.ac.jp},
Ichi Tanaka$^{2}$,
Masaru Kajisawa$^{1}$,
Jaron Kurk$^{3}$,
Bram Venemans$^{4}$,
Carlos De Breuck$^{5}$,
Jo\"el Vernet$^{5}$,
Chris Lidman$^{6}$
}
\vspace*{6pt}\\
$^{1}$National Astronomical Observatory of Japan, Mitaka, Tokyo 181--8588, Japan\\
$^{2}$Subaru Telescope, National Astronomical Observatory of Japan, 650 North Aohoku Place, Hilo, HI 96720, USA\\
$^{3}$Max-Planck-Institute f\"ur Astronomie, K\"onigstuhl 17, 69117 Heidelberg, Germany\\
$^{4}$Institute of Astronomy, University of Cambridge, Madingley Road, Cambridge CB3 0HA, UK\\
$^{5}$European Southern Observatory, Karl-Schwarzschild-Str. 2, D-85748 Garching, Germany\\
$^{6}$European Southern Observatory, Alonso de Cordova 3107, Vitacura, Casilla 19001, Santiago 19, Chile
}

\begin{document}

\maketitle

%
%
\begin{abstract}
We explore the evolved galaxy population in the proto-clusters around four high-$z$
radio galaxies at $2\lsim z\lsim3$ based on wide-field near-infrared (NIR) imaging.
Three of the four fields are known proto-clusters 
as demonstrated by over-densities of line emitting galaxies
at the same redshifts as the radio galaxies found by narrow-band
surveys and spectroscopic follow-up observations.
We imaged the fields of
three targets (PKS~1138--262, USS~0943--242 and MRC~0316--257) 
to a depth of $K_s\sim22$ (Vega magnitude, 5$\sigma$) over a $4'\times7'$ area 
centered on the radio galaxies with
a new wide-field NIR camera, MOIRCS, on the Subaru Telescope.
Another target (USS~1558--003) was observed with SOFI on the NTT to a depth
of $K_s=20.5$ (5$\sigma$) over a $5'\times5'$ area.
We apply colour cuts in $J-K_s$ and/or $JHK_s$ in order to exclusively
search for galaxies located at high redshifts: $z>2$.
To the 5$\sigma$ limiting magnitudes, we see a significant excess of NIR selected galaxies
by a factor of two to three compared to those found in the field of {\it GOODS-South}.
The spatial distribution of these NIR selected galaxies is not uniform and traces 
structures similar to those of emission line galaxies, although 
the samples of NIR selected galaxies and emitters show little overlap,
from which we conclude that the former tend to be an evolved population with much 
higher stellar mass than the latter, young and active emitters.
We focus on the NIR colour--magnitude sequence of the evolved population and
find that the bright-end (M$_{\rm stars}>10^{11}$\msun) of the red sequence
is well populated by $z\sim2$ but much less so
in the $z\sim3$ proto-clusters.  This may imply that the bright-end of the
colour--magnitude sequence first appeared between $z=3$ and 2,
an era coinciding with the appearance of submm galaxies
and the peak of the cosmic star formation rate.
Our observations show that during the same epoch, massive
galaxies are forming in high density environments by vigorous star
formation and assembly.
\end{abstract}

\begin{keywords}
galaxies: clusters: general ---
galaxies: formation: ---
galaxies: evolution: ---
galaxies: high-redshift 
\end{keywords}

%
%
\section{Introduction}

\label{sec:intro}

The colour--magnitude relation is a well established scaling relation
seen in cluster early-type galaxies locally (e.g., Visvanathan \& Sandage 1977,
\citealp{bow92}), and is well recognised even in distant clusters at least
out to $z\sim1$ (e.g., \citealp{ell97}; \citealp{kod98}; \citealp{sta98};
\citealp{van98}; 2001; Blakeslee et al.\ 2003; Stanford et al.\ 2006).
The tightness of the relation and
its colour evolution with redshift
indicate that the early-type galaxies are dominated by
old stellar populations formed before a redshift of two.
The mass assembly of cluster early-type galaxies is also expected
to take place early ($z>1$) by the lack of evolution in their stellar mass
functions to $z\sim1$
(e.g., Kodama et al. 2003; Ellis \& Jones 2004; Strazzullo et al.\ 2006).
Therefore, the massive-end of the relation should have been established
since early times ($z\gg 1$).
Identifying the epoch of massive galaxy formation is important since it
can place strong constraints on the bottom-up formation of galaxies
predicted by theories of the CDM Universe (e.g., Kauffmann \& Charlot 1998,
De Lucia et al. 2006).
By looking back further in time, we should eventually be able to witness
the first appearance of the bright end of the red sequence galaxies in
proto-clusters, and hence we can observationally identify the formation
epoch of massive galaxies.

Despite its obvious immediate importance, however, such a study has been
largely limited until very recently by the following two factors.
First of all, samples of proto-clusters were very sparse not only due to
the observational difficulty in finding them due to the faintness of the
targets but also due to the intrinsic underdevelopment of the structures
at high redshifts.
Secondly, the small field of view of near-infrared (NIR) cameras has
prevented us from sampling the rest-frame optical light that provides
most of the information on the more evolved
stellar population, for a large number of high-$z$ galaxies.
A large amount of work has been devoted to surveys for line emitters around high redshift
radio galaxies using narrow-band imaging and/or spectroscopy in the optical (e.g., \citealp{kur00},
\citealp{ven02}; \citealp{ven07}; Steidel et al.\ 2000; 2005), which in fact successfully
found a number of over-dense regions at high redshifts traced by young star forming galaxies.
The Lyman break technique based on the broad-band multi-colour imaging in the optical
has also been used to survey some concentrations of high-$z$ galaxies such as at $z\sim3$
(e.g., Steidel et al.\ 1998).

The advent of a new wide-field NIR camera MOIRCS
(Multi-Object Infra-Red Camera and Spectrograph; Ichikawa et al.\ 2006)
on Subaru (providing a $4'\times7'$ field of view) has extended our ability
of searching for proto-clusters in the NIR.
In this paper, we present photometric properties and spatial distribution
of galaxies in the fields of four known proto-clusters around radio galaxies
at $2\lsim z\lsim3$.
The candidate proto-cluster member galaxies associated to the central radio
galaxies are extracted on the basis of their NIR colours.
We will focus on the appearance of the massive end of the red sequence
of galaxies in the proto-clusters.

The structure of the paper is the following.  Our observational data-set
will be described in \S2, and the colour selection of proto-cluster member
candidates will be explained in \S3.  The results on the colour-magnitude diagrams
and the spatial distribution will be given in \S4 and \S5, respectively and a summary
is given in \S6.  The Vega-referred magnitude system and cosmological parameters 
of H$_0$=70~km s$^{-1}$ Mpc$^{-1}$, $\Omega_{m}=0.3$, and
$\Omega_{\Lambda}=0.7$ are used throughout the paper.

%
%
\section{Targets, Observation and Data Reduction}
\label{sec:obs}

\subsection{Target selection}

Radio galaxies are often used as markers of high
density regions at high redshifts, since they are among the most
massive galaxies at any redshift
($M_{\rm star}>10^{11} M_{\odot}$, \citealp{roc04}).
In fact, about half of the powerful radio galaxies at intermediate
redshift inhabit rich environments (\citealp{hil91}, \citealp{bor06}).
Also, given the correlation between AGN activity and bulge mass
(Magorrian relation, \citealp{geb00}), it is natural to imagine that massive
cD galaxies sitting in deep potential wells of clusters tend to host massive
powerful AGN which are therefore identified
as radio galaxies at high redshifts.
Narrow-band surveys of Ly$\alpha$ emitters around high-$z$ radio galaxies
up to $z\sim5$ have been intensively conducted by the Leiden group (e.g., \citealp{kur00},
\citealp{ven02}, \citealp{ven07}) over many years revealing a large number of emitters
at the redshifts of the radio galaxies, and a large fraction of those
emitters have been spectroscopically confirmed.
The over-dense regions traced by the star forming galaxies are embedded in large scale
structures in the early Universe and are likely to evolve into massive systems
at the present day, such as clusters of galaxies.
These are therefore ideal sites to study
the ancestors of present-day early-type galaxies in their formation phase.

Recently, some NIR surveys have been carried out searching for
``evolved'' galaxy populations around radio galaxies at $1\lsim z\lsim 3$,
an approach complementary to the line emitter surveys.
Some over-dense regions of red galaxies were reported, which are probably associated
to the radio galaxies
(e.g., \citealp{bes03}, \citealp{hal98}, \citealp{kod03}, \citealp{wol03}, \citealp{tof03}).
Kajisawa et al.\ (2006) have recently conducted deep $JHK'$ surveys with
CISCO on Subaru, and found two convincing proto-clusters at $z\sim2.5$
that show over-densities of NIR-selected galaxies by more than a factor of three
compared to the general field {\it GOODS-South} (hereafter {\it GOODS-S}).

From the currently available sample of proto-cluster candidates,
we select four targets, of which three
(PKS~1138--262, USS~0943--242 and MRC~0316--257)
are fields with a number of spectroscopically
confirmed \lya\ and/or \ha\ emitters associated with the radio galaxies.
Another target (USS~1558--003) is selected from the sample of \cite{kaj06}
as, in earlier CISCO observations, it shows a clear over-density of NIR-selected
galaxies. No line emitter survey has been conducted around this target yet.
All the targets lie at $2\lsim z\lsim3$ and are listed in Table~\ref{tab:obs}.
Hereafter, they are sometimes referred to as 1138, 0943, 0316 and 1558 for short.

\subsection{Observation and data reduction}

\begin{table*}
 \centering
  \caption{Summary of the observational data}
  \label{tab:obs}
  \begin{tabular}{@{}lcccccccc@{}}
  \hline\hline
   Name        & Redshift & Telescope/    & \multicolumn{3}{c|}{Exp. Time (min)} & PSF FWHM & FoV & lim. mag.\\
               &          & Instrument    & $J$ & $H$ & $K_s$ &(arcsec)& (arcmin$^2$) & (5$\sigma$) \\ 
 \hline
PKS~1138$-$262 & 2.156    & Subaru/MOIRCS &  83 & --- &  55 & 0.5--0.7 & 25 & $J$=23.3, $K_s$=22.0\\ 
USS~1558$-$003 & 2.527    & NTT/SOFI      & 180 & --- & 175 & 0.6--0.7 & 25 & $J$=22.5, $K_s$=20.5\\ 
USS~0943$-$242 & 2.923    & Subaru/MOIRCS & 118 &  68 &  63 & 0.4--0.6 & 25 & $J$=23.5, $H$=22.6, $K_s$=22.0\\ 
MRC~0316$-$257 & 3.130    & Subaru/MOIRCS &  78 &  60 &  55 & 0.6--0.7 & 25 & $J$=23.0, $H$=22.6, $K_s$=21.8\\ 
\hline
\end{tabular}
\end{table*}

We performed NIR imaging observations in the $J$, $H$ and $K_s$ bands 
of the fields of three high-$z$ radio galaxies at $2\lsim z\lsim3$ with MOIRCS on the
Subaru Telescope on 2006 January 6--7.
MOIRCS has a field of view (FoV) of 4 $\times$ 7 arcmin$^2$ with a 0.117$''$
pixel scale covered by two 2k by 2k CCDs, which provide a co-moving area
of 6 $\times$ 10.5~Mpc$^2$ at $z\sim2$ and 7.5 $\times$ 13~Mpc$^2$ at 
$z\sim3$.  Two corners of the MOIRCS CCDs are vignetted.
After removing these,
the final areal coverage used is about 25~arcmin$^2$.
Our targets and the on-source exposure times are given in Table \ref{tab:obs}.
The weather conditions were stable during the observations, and the seeing
varied between 0.4$''$ and 0.7$''$ (FWHM) for most of the nights. 

The fourth radio galaxy field, USS1558--003, was imaged with SOFI on the 
NTT in the $J$ and $K_s$ bands on 2006 March 20-22 under photometric and
good seeing conditions. The pixel scale of SOFI is 0.288$''$ and its FoV is 
4.9$\times$4.9 arcmin$^2$, close to the effective MOIRCS FoV.

The MOIRCS data were reduced using a purpose-made pipeline software
package called {\tt MCSRED} (MoirCS REDuction; Tanaka et al.\ 2007).
We performed flat-fielding with the super-flat frames constructed by combining
our object frames without being registered, and subtracted dark and sky background.
Then the data were co-registered and combined.  The SOFI data were reduced
in the same manner.

For each target, the images were convolved with a Gaussian kernel 
to match the PSF of the image taken under the worst seeing
(see Table~\ref{tab:obs}).

Source detection was performed in the $K_s$-band images using the
SExtractor image analysis package \citep{ber96}. 
We adopted MAG\_AUTO from SExtractor as the total $K_s$-band
magnitude of detected objects.  For colour measurements in $J-K_s$ and $H-K_s$, 
we used aperture magnitudes MAG\_APER with a 1.5$''$ diameter aperture,
using the dual image mode with detection in the $K_s$-band.
The UKIRT Faint Standards were used for the flux calibration.
The 5$\sigma$ limiting magnitudes for each field are given in Table~\ref{tab:obs}.
All magnitudes are corrected for Galactic extinction, which is
estimated at the positions of the radio galaxies (near the center of each
observed field) based on \cite{sch98}.

%
%
\section{Colour selection of proto-cluster member candidates}
\label{sec:colsel}

\begin{figure*}
\begin{center}
\leavevmode
  \epsfxsize 0.47\hsize
  \epsfbox{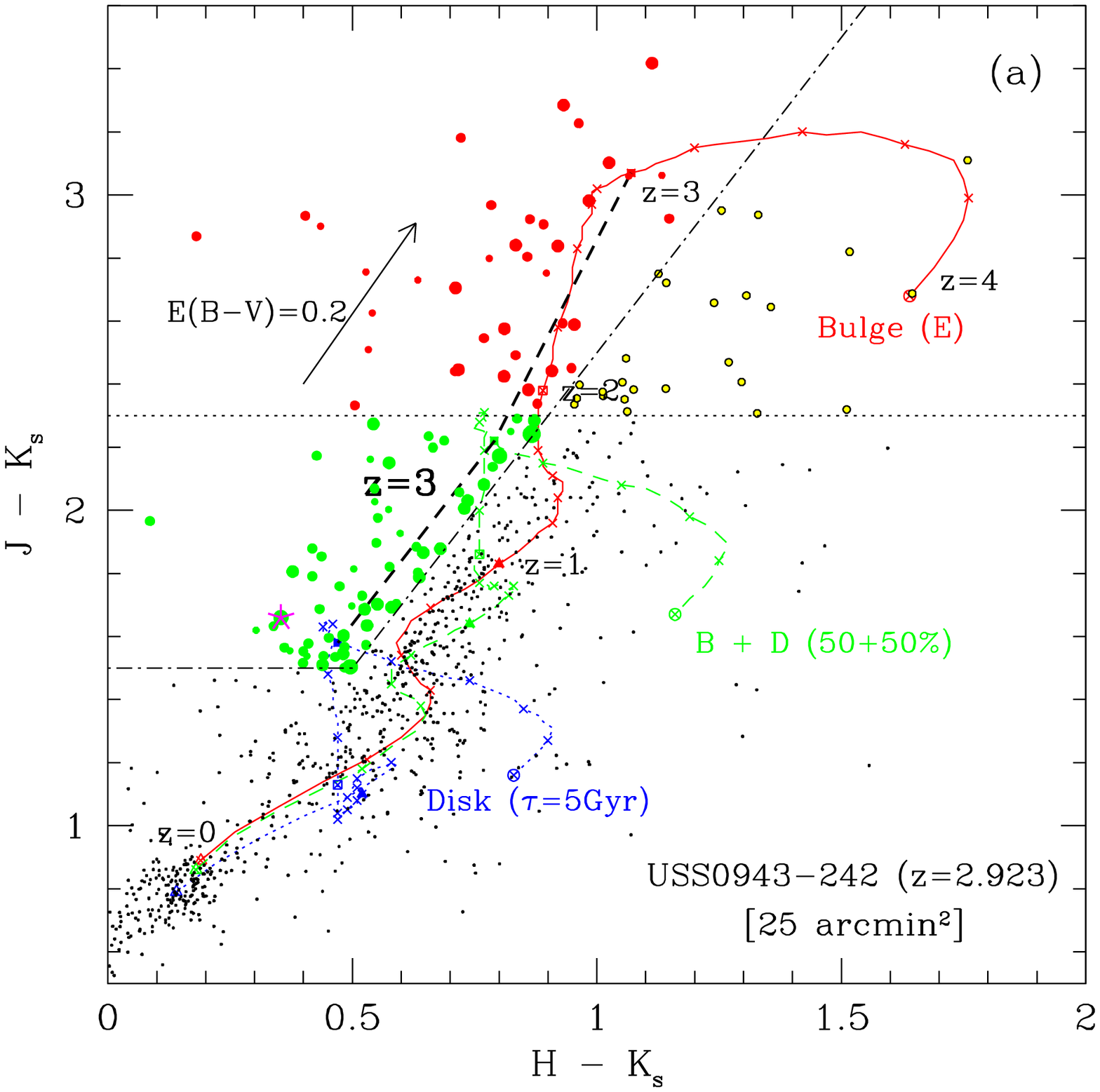}
  \epsfxsize 0.47\hsize
  \epsfbox{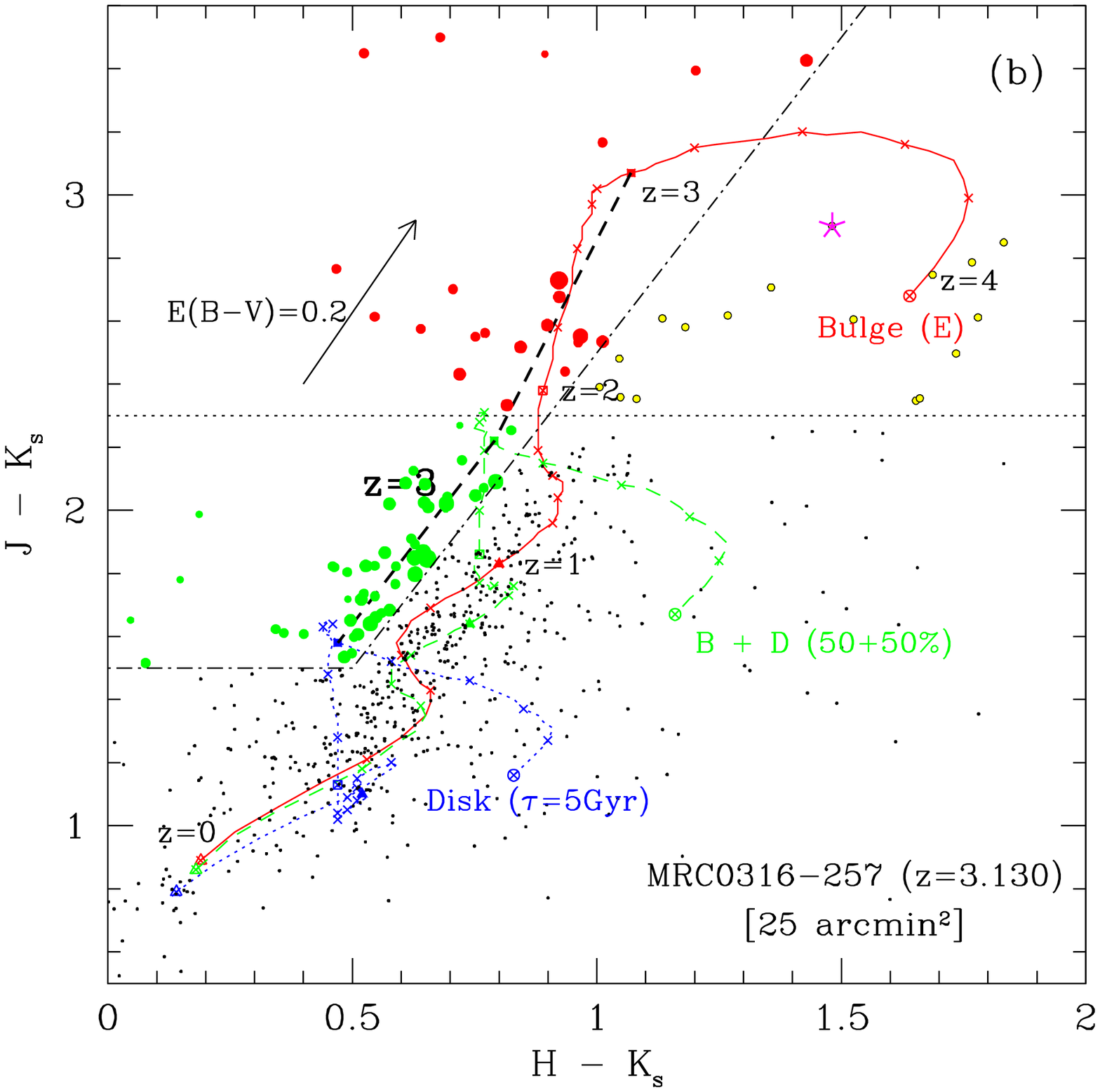}
\end{center}
\caption{
Colour--colour diagrams of the two proto-clusters USS~0943--242 at $z\sim2.9$
(left panel) and MRC~0316--257 at $z\sim3.1$ (right panel).
Galaxies down to 5$\sigma$ limiting magnitudes in $K_s$ are plotted
($K_s$=22 for 0943 and $K_s$=21.8 for 0316).
The horizontal dotted line and the dot-dashed lines show the
boundaries of our single colour selection of $J-K_s>2.3$ and the
two-colours selection with $JHK_s$, respectively.
Solid, dashed and dotted curves represent the evolutionary tracks
of galaxies over $0<z<4$ with different star formation histories
(passive, intermediate, and active) formed at $z_{\rm form}=5$ \citep{kod99}.
The thick dashed line connects the model points at $z=3$, showing
that our selection technique works to pick out not only passively evolving
galaxies at $2\lsim z\lsim3.2$ but also star forming galaxies at $2.4\lsim z\lsim3.1$.
Red filled circles and green filled circles are r-JHK and b-JHK populations, respectively.
The size of the symbols is scaled according to apparent magnitudes in $K_s$-band
(where larger means brighter).
Yellow filled circles with black boundaries show the galaxies that satisfy
the DRG criterion but not the JHK criterion.
A large star indicates the targeted radio galaxies.
The arrow shows a reddening vector of Calzetti et al.'s law \citep{cal00} of 
E$(B-V)=0.2$ at $z=3$.  Importantly the vector is almost parallel
to the boundary line of our JHK selection, and thus our method is robust
against dust extinction.
Moreover, contamination from lower-$z$ ($z\lsim2$) galaxies due to dust extinction
is expected to be minimal. Note that the direction of reddening vector on this diagram
hardly changes for $0<z<3$ (less than 12 per cent in the slope).
}
\label{fig:jhk}
\end{figure*}

\begin{figure*}
\begin{center}
\leavevmode
  \epsfxsize 0.47\hsize
  \epsfbox{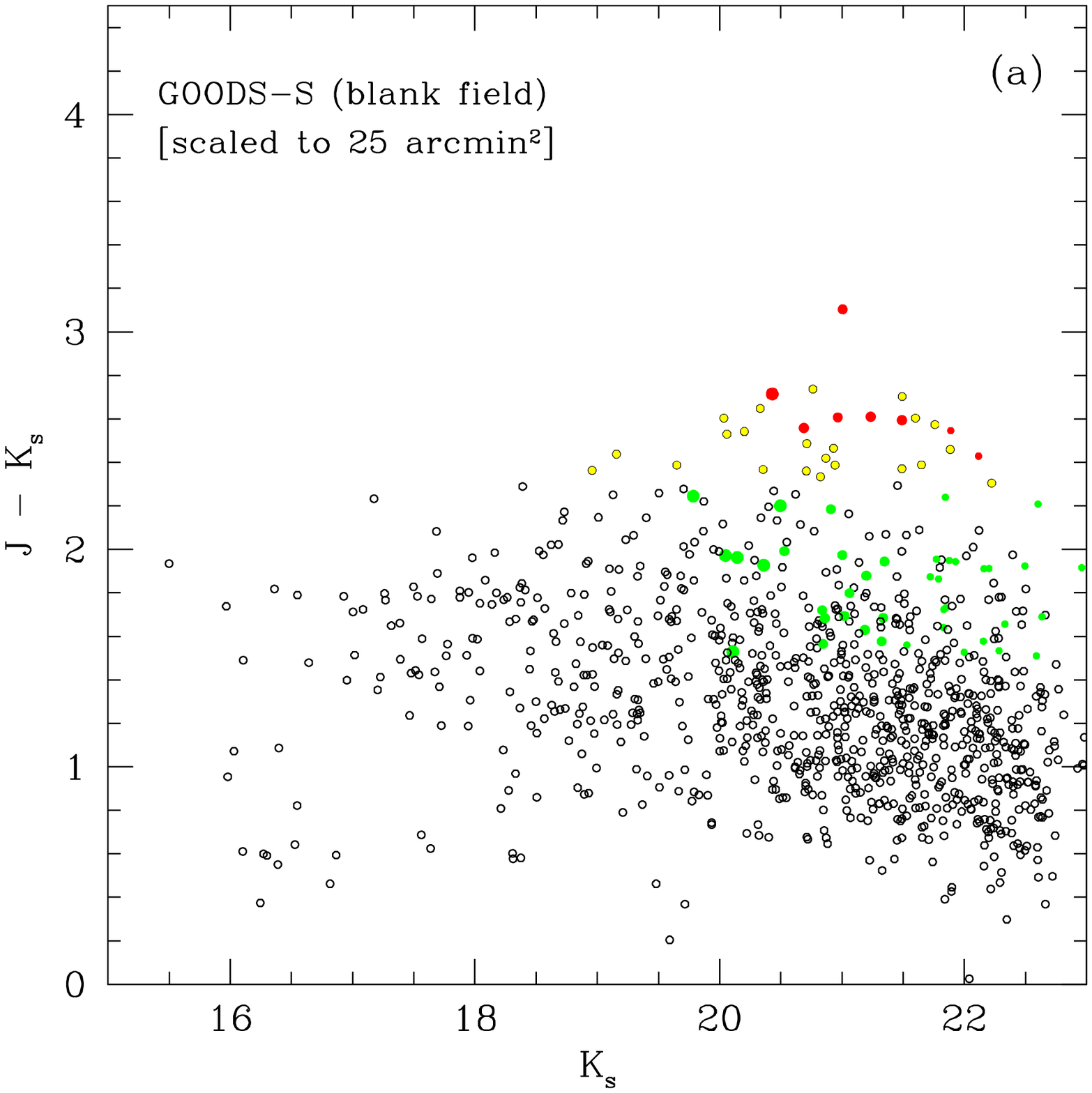}
  \epsfxsize 0.47\hsize
  \epsfbox{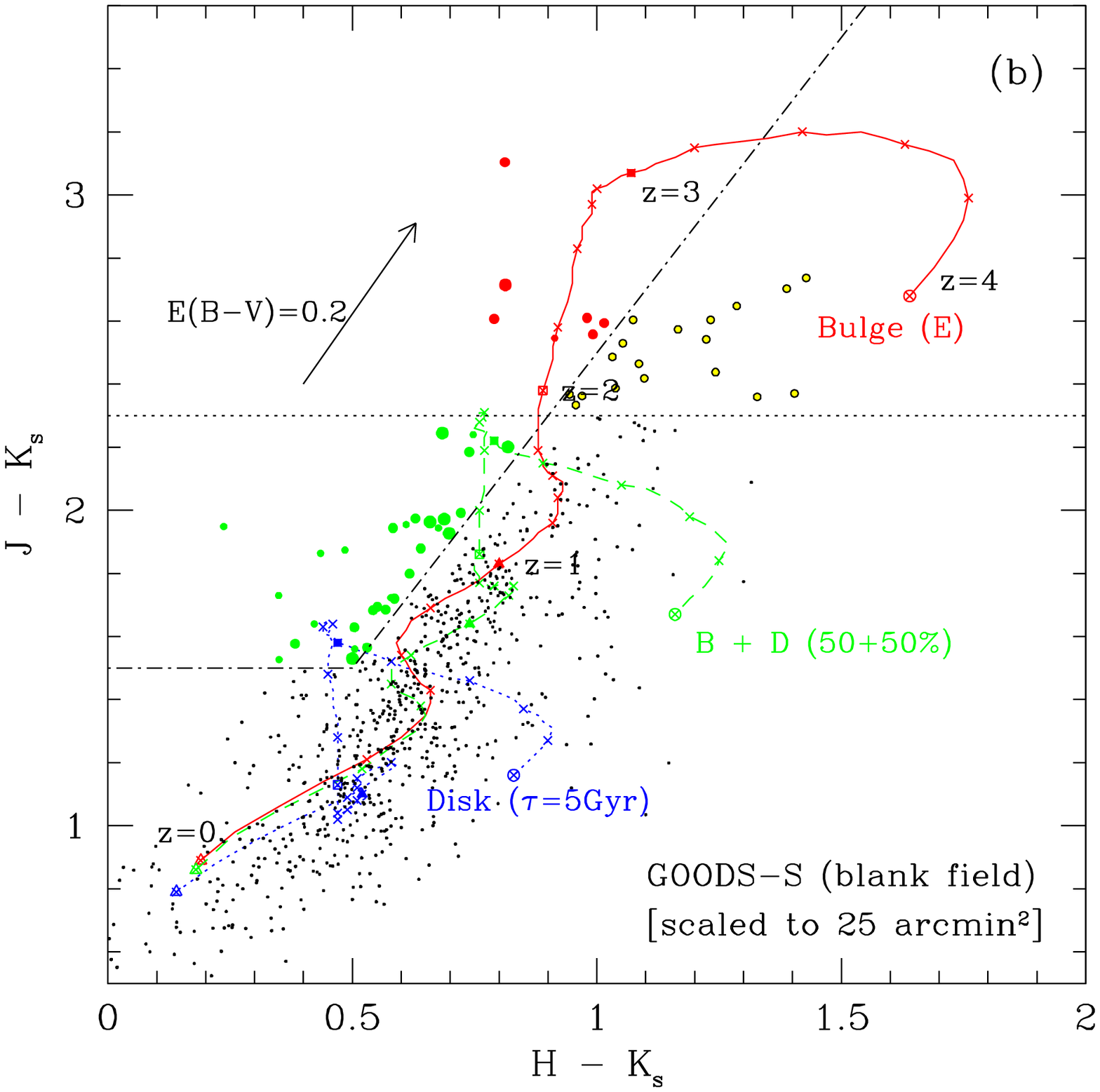}
\end{center}
\caption{
A colour--magnitude (left) and a colour--colour (right) diagrams of
the {\it GOODS-S} field \citep{gia04}, scaled down by randomly sampling the galaxies 
to 25 arcmin$^2$ in order to match the FoV of the radio galaxies fields.
In the colour--colour diagram, only the galaxies brighter than $K_s=22$
are shown.
These plots can therefore be directly compared with the proto-cluster
regions shown in Figs.~\ref{fig:jk} and \ref{fig:jhk}.
See those figures for meanings of the symbols.
}
\label{fig:goodss}
\end{figure*}

%
%
%
\subsection{A new definition of NIR-selected galaxies at $2\lsim z\lsim3$}

Our original $K_s$-selected catalogues contain $\sim$700--1000 objects
per field and most of them should be foreground/background galaxies or
galactic stars which are not physically associated to the
proto-clusters around the targeted radio galaxies.  It is therefore
essential to remove those contaminations as much as possible while
keeping most of the member candidates in order to examine the galaxy
population in the proto-clusters.

For this purpose, we apply colour selections either by $J-K_s>2.3$ or
in $JHK_s$ colour--colour space.  The former criterion is a single
colour cut in $J-K_s$, devised by the FIRES team
(Faint InfraRed Extragalactic Survey; Franx et al.\ 2003).
The galaxies that satisfy this criterion were named Distant Red
Galaxies (DRGs) and they have been shown to correspond to $2<z<3$
galaxies with a high completeness rate when followed up
spectroscopically (e.g., \citealp{van03}, \citealp{for04},
\citealp{red05}).  In fact, such red colours of galaxies can only be
reproduced either by Balmer/4000\AA-break galaxies with old
populations at $z\gsim2$ or foreground galaxies with heavy dust
extinction (Fig.~\ref{fig:jhk}).  The latter criterion is based
on cuts in the three $JHK_s$ bands, devised by some of the authors,
first defined and applied in \cite{kaj06}.
Figure~\ref{fig:jhk}
illustrates the $J-K_s$ versus $H-K_s$ colour--colour diagrams of our
two $z\sim3$ targets.  As shown by three model tracks and the thick
dashed line connecting the $z=3$ points, most of the galaxies at
$2.4\lsim z\lsim3.1$ are expected to lie at the top left corners of
these diagrams.  Therefore, if we apply the colour cuts shown by the
dot-dashed lines (ie., $J-K_s>2\times(H-K_s)+0.5\ \&\ J-K_s>1.5$), we
can effectively isolate these galaxies.
This criterion is met while the Balmer/4000\AA\-break of galaxies
falls between $J$-band and $H$-band at this redshift range.
In this paper, we further classify the JHK-selected galaxies
(hereafter JHKs)
into red and blue populations separated at $J-K_s=2.3$, and we hereafter
call them r-JHK and b-JHK, respectively.  The definitions of the galaxy classes are
summarised in Table~\ref{tab:def}.

This two-colour-based selection in $JHK_s$ has a significant advantage
over the classical DRG selection in the single $J-K_s$ colour, since
the two-colour cut can pick out not only passive or dusty galaxies but
also younger or star forming galaxies at $2.5\lsim z\lsim3$ which have
relatively bluer colours in $J-K_s$ and would have been missed by the
single $J-K_s>2.3$ cut.  It should be also noted that this selection
is robust against dust extinction, since the reddening vector
(indicated by an arrow in Fig.~\ref{fig:jhk}) is almost parallel to
the boundary line of the JHK-selection.
At the same time, contamination from lower-$z$ galaxies ($z\lsim2$)
due to dust extinction is expected to be minimal.
In addition, cool M, L and T dwarfs are also efficiently excluded by this
criterion, as they are either too blue in $J-K_s$ or located on the
right side of the boundary line of the JHK-selection \citep{kaj06}.
By using this JHK-selection technique, \cite{kaj06} has identified
proto-cluster candidates with evolved galaxy populations around some
radio galaxies at $z\sim2.5$.  It is notable that 5 out of 6 radio
galaxies themselves (which we know are located at $z\sim2.5$ by
spectroscopy) satisfied our JHK colour selection criteria.  This means
that our selection actually works for blue members with on-going star
formation even though the strong emission lines from an AGN component
could affect their colours \citep{iwa03}.  The central radio galaxies
0943 and 0316 are labeled by stars in Fig.~\ref{fig:jhk}, which shows
that 0943 satisfies the b-JHK criterion while 0316 does not.
Furthermore, galaxies that are selected by our JHK cut show a clear
statistical excess
(factor 2$\sim$4)
in two of the fields when they are
compared to the general field of {\it GOODS-S}.  This is true for the
bluer JHK population (b-JHK) as well, which further supports the
effectiveness of our selection technique.

\begin{table}
\caption{Definitions of various NIR selected galaxies}
\label{tab:def}
\begin{tabular}{ll}
\hline\hline
category & definition \\
\hline
DRG    & $J-K_s>2.3$ \\
\hline
JHK    & $J-K_s>2\times(H-K_s)+0.5\ \&\&\ J-K_s>1.5$ \\
 \hspace*{0.4cm} r-JHK  & JHK\ \ \&\&\ \ $J-K_s>2.3$ \\
 \hspace*{0.4cm} b-JHK  & JHK\ \ \&\&\ \ $J-K_s<2.3$ \\
\hline
\end{tabular}
\end{table}

We apply the same JHK-selection for the two proto-clusters at $z\sim3$
(0943, 0316) in this paper for which a complete $JHK_s$ data-set is
available.  For the remaining two targets at $2\lsim z\lsim2.5$ (1138,
1558), we apply the single colour cut of $J-K_s>2.3$ instead.

The NIR selected galaxies are highlighted in Fig.~\ref{fig:jhk} with
different symbols.  Red filled circles and green filled circles are
r-JHK and b-JHK populations, respectively.  The size of the symbols is
scaled according to the $K_s$-band magnitude (bigger ones are
brighter).  Yellow filled circles with black boundaries show the
galaxies that satisfy the DRG criterion but not the JHK criterion, and
hence are possibly background objects.  All other objects detected in
the $K_s$ band are shown by small dots.

\subsection{Over-density of the NIR-selected galaxies}
\label{sec:overdensity}

\begin{figure*}
\begin{center}
\leavevmode
  \epsfxsize 0.47\hsize
  \epsfbox{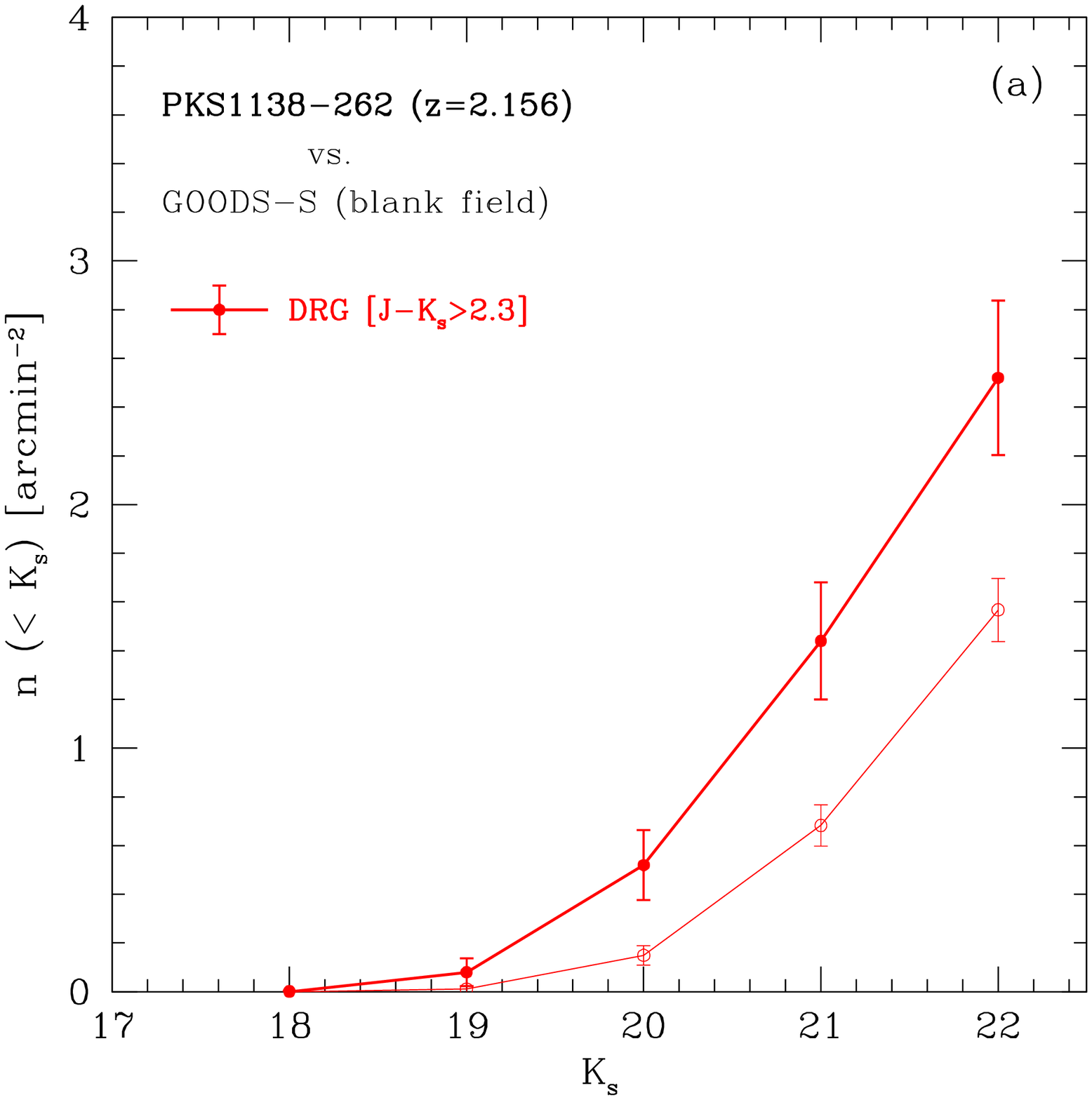}
  \epsfxsize 0.47\hsize
  \epsfbox{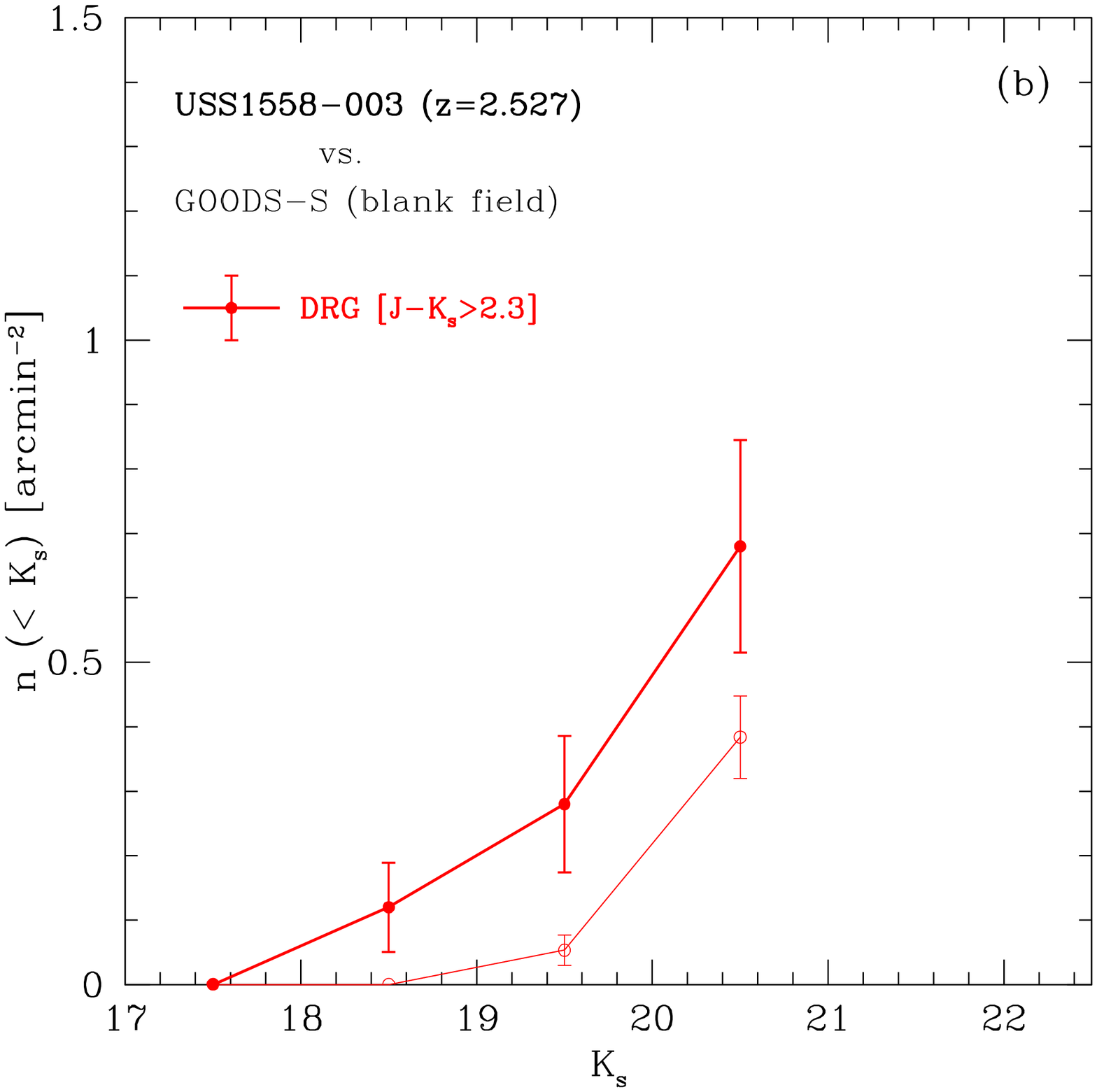}\\
  \hspace*{-0.1cm}
  \epsfxsize 0.47\hsize
  \epsfbox{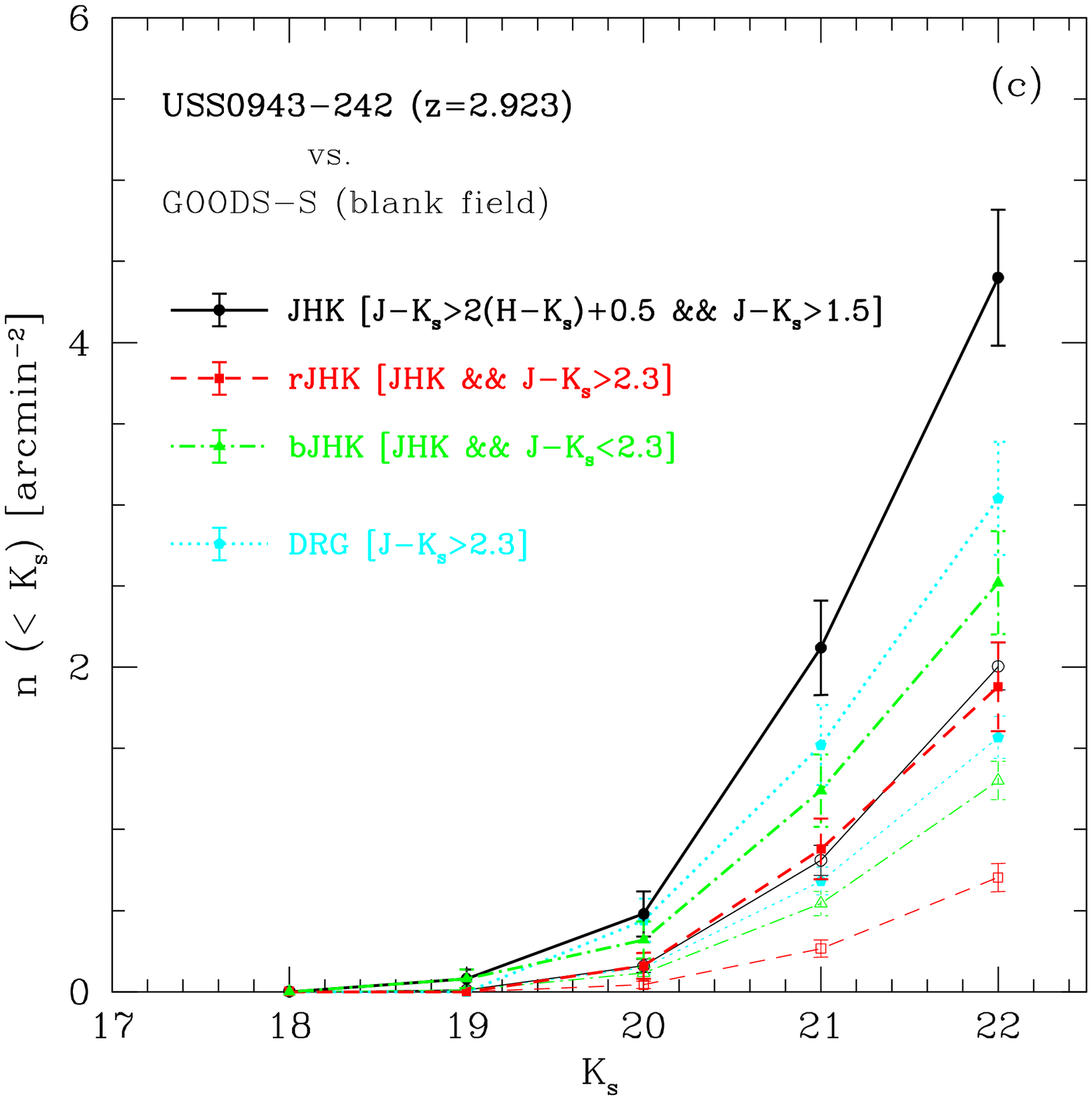}
  \epsfxsize 0.47\hsize
  \epsfbox{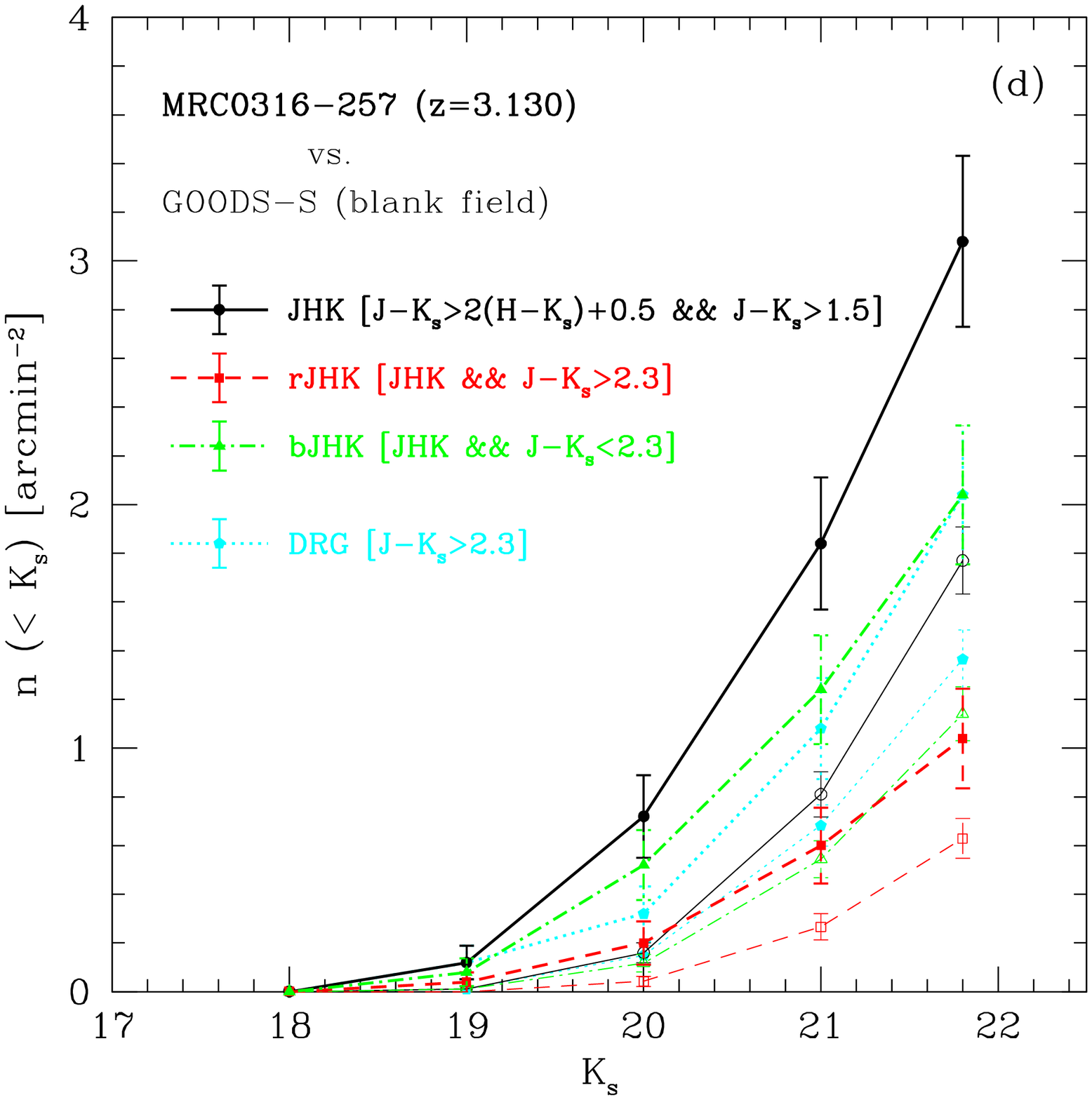}
\end{center}
\caption{
Combined cumulative number counts of galaxies in the radio galaxies
fields (thick lines) and those in a control field {\it GOODS-S} (thin lines).
Error-bars are based on Poisson statistics.
Data for the full effective area of 94~arcmin$^2$ are used for the counts in {\it GOODS-S}
\citep{gia04}.
}
\label{fig:count}
\end{figure*}

Fig.~\ref{fig:jhk} shows that there are
many ($>70$)
galaxies which satisfy the JHK selection in each of the two $z\sim3$ fields. 
To assess the over-density of the NIR selected galaxies compared to the
general field, we use the public VLT/ISAAC version~1.5 data of the
{\it GOODS-S} field
(Vandame et al., in prep; see also Giavalisco et al. 2004),
The latter data are reproduced in
Fig.~\ref{fig:goodss} on NIR colour--magnitude and colour--colour
diagrams.  For a direct comparison with our proto-cluster fields, the
effective areal coverage of the {\it GOODS-S} data was scaled down from
94~arcmin$^{2}$ to 25~arcmin$^{2}$ by randomly sampling galaxies from
the catalog.  By comparing Fig.~\ref{fig:jhk} with Fig.~\ref{fig:goodss},
it is obvious that the JHK-selected galaxies in the two proto-cluster
regions at $z\sim3$ are a lot more numerous than in the scaled {\it
GOODS-S} area.

To present this more quantitatively, we show the cumulative number
counts of the NIR selected galaxies in the proto-cluster fields and in
the {\it GOODS-S} field in Fig.~\ref{fig:count}.  For 1138 and 1558
only DRGs are plotted, while for 0943 and 0316 JHKs (and its
subsamples r-JHK and b-JHK) as well as DRGs are plotted.  We find
clear statistical excesses in all classes of NIR selected galaxies in
all of the four proto-cluster fields (thick curves) compared to {\it
GOODS-S} (thin curves).  The excess factors range from 1.5 to 4
depending on the field, galaxy class, and depth considered.
In particular, the DRGs in 1138 and 1558 at $K_s<20$, and JHKs in 0943
and 0316 at $K_s<21$ show the largest excesses.
These excess factors of the NIR-selected galaxies are consistent with
those of the surface number densities of the \lya\, emitters
which are factor 2, 3 and 3 for 1138, 0943 and 0316, respectively
(Kurk et al.\ 2004a; Venemans et al.\ 2007).
To the 5$\sigma$
limiting magnitudes of each target (see Table~\ref{tab:obs}), the
excess numbers of the NIR-selected objects are: 23 and 8 DRGs in 1138 and
1558 respectively, and 60 and 32 JHKs in 0943 and 0316 respectively.
An indication for the existence of such red galaxies observed as EROs
(Extremely Red Objects) was also found in Kurk et al.\ (2004a).

This strongly suggests that at least some of the NIR-selected galaxies
are physically clustered and associated to the central radio galaxies.
Because the $K_s$-bright galaxies ($K_s<21$) are relatively massive
($M_{\rm stars}>2\times10^{10}$ M$_\odot$) as we discuss later, these
proto-clusters already seem to contain an evolved galaxy population as
well as young \lya/\ha\ emitters (Kurk et al.\ 2004a; 2004b; Venemans
et al.\ 2007) which tend to be much less massive, typically by more
than an order of magnitude (Gawiser et al.\ 2006; Venemans et al.\
2005).  It is likely that these fields will evolve into rich clusters
of galaxies, today dominated by old passively evolving galaxies.

There is a caution, however, that field-to-field variation in the number
density of intrinsically rare massive galaxies is not negligible.
In fact, the {\it GOODS-S} field that we use for comparison could be
a slightly under-dense region of massive galaxies as suggested by
van Dokkum et al.\ (2006).  They claim that the number density of massive
galaxies ($>$10$^{11}$\msun) at $2<z<3$ in the {\it GOODS-S} field is
$\sim$ 60\% of the average when compared to the {\it MUSYC} survey.
Since we have computed over-densities using the {\it GOODS-S} field,
our numbers may be over-estimated by at most a factor of two if compared
with a larger (less biased) blank field.
Therefore the excess factors of the NIR-selected galaxies could be reduced
to 1.5$\sim$2 at the bright-end ($K_s$$<$19.5--20) in such a case.

%
%
\section{The first appearance of massive red sequence galaxies}
\label{sec:cm}

\begin{figure*}
\begin{center}
\leavevmode
  \epsfxsize 0.47\hsize
  \epsfbox{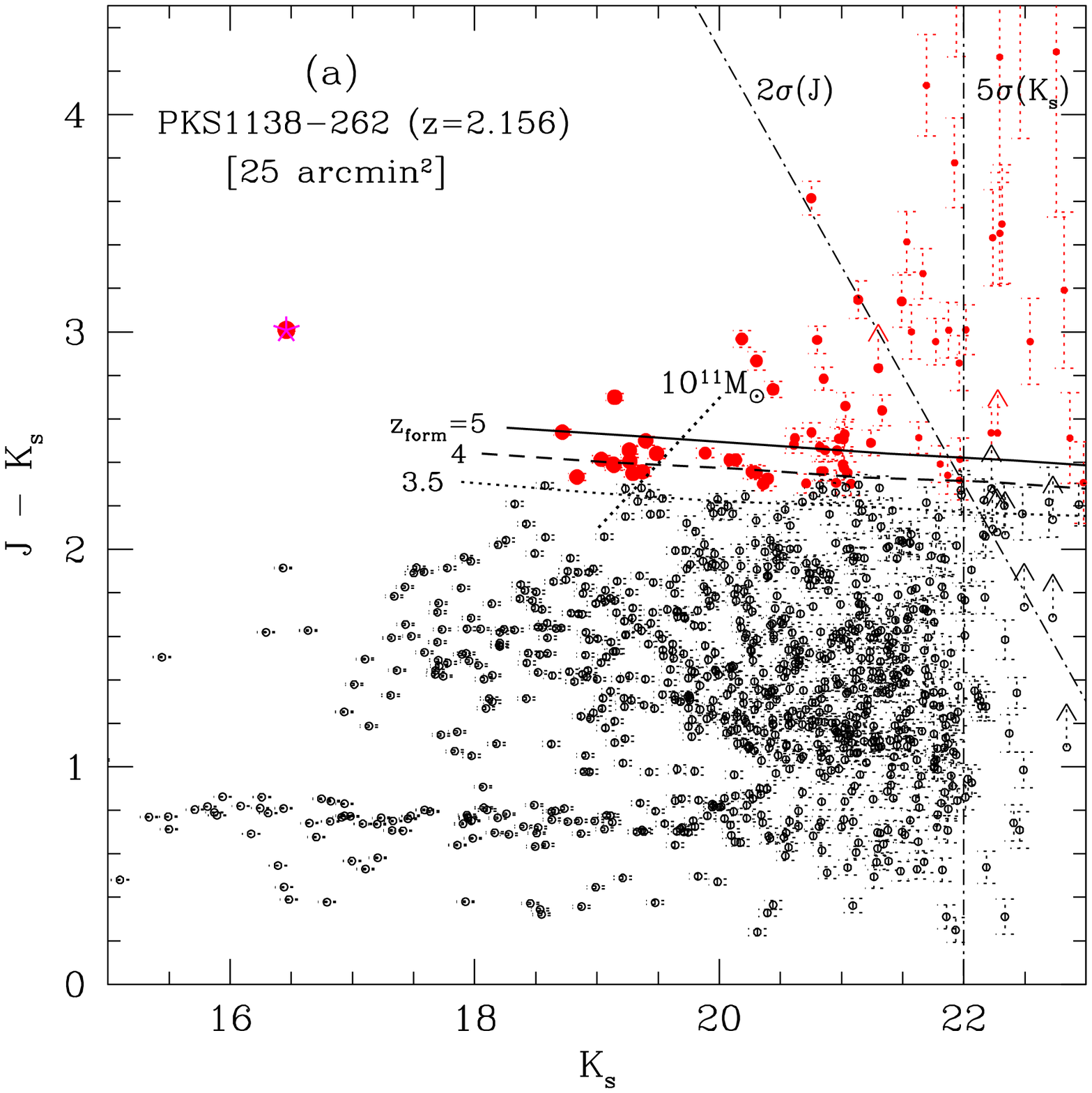}
  \epsfxsize 0.47\hsize
  \epsfbox{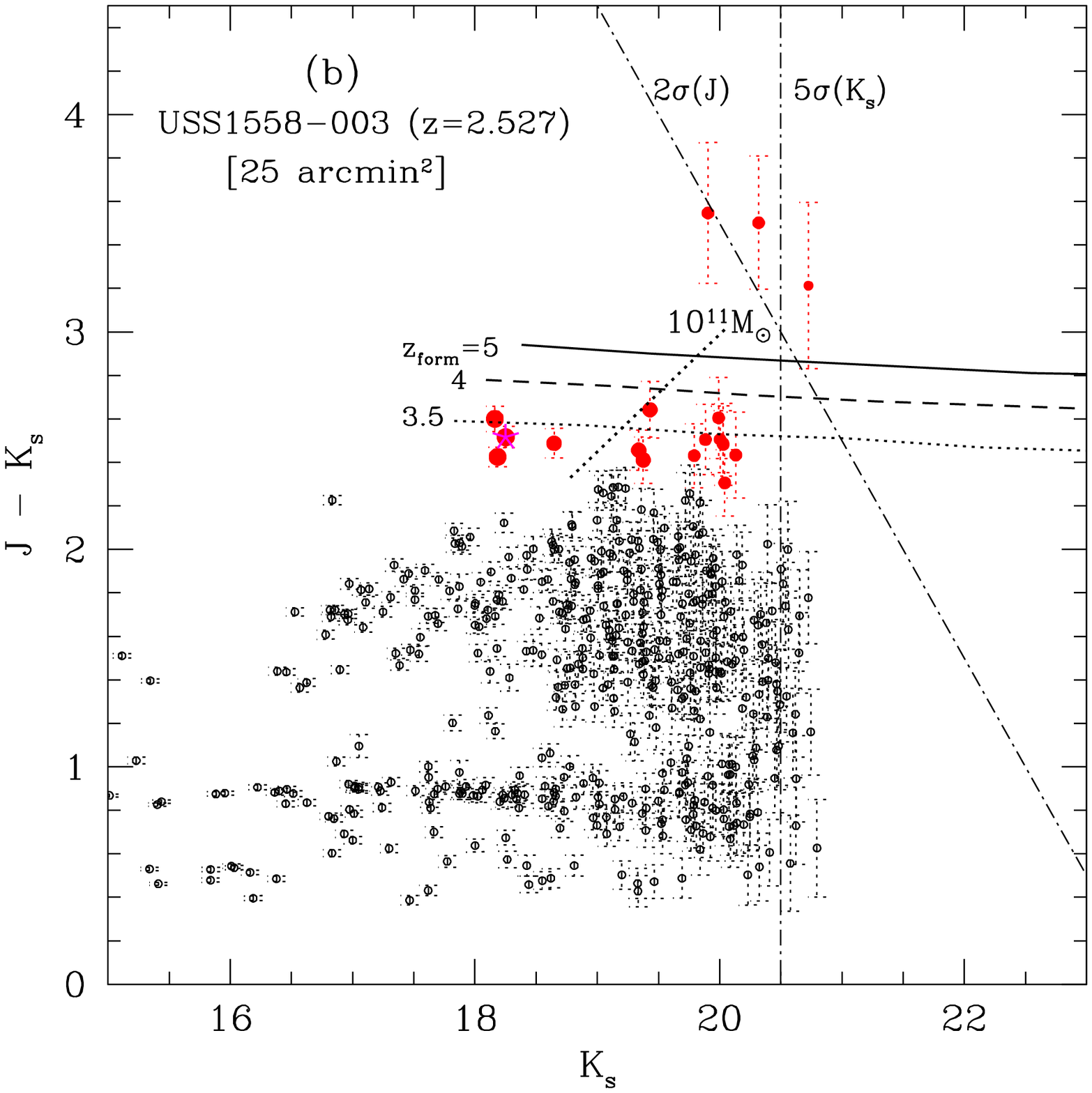}\\
  \hspace*{-0.1cm}
  \epsfxsize 0.47\hsize
  \epsfbox{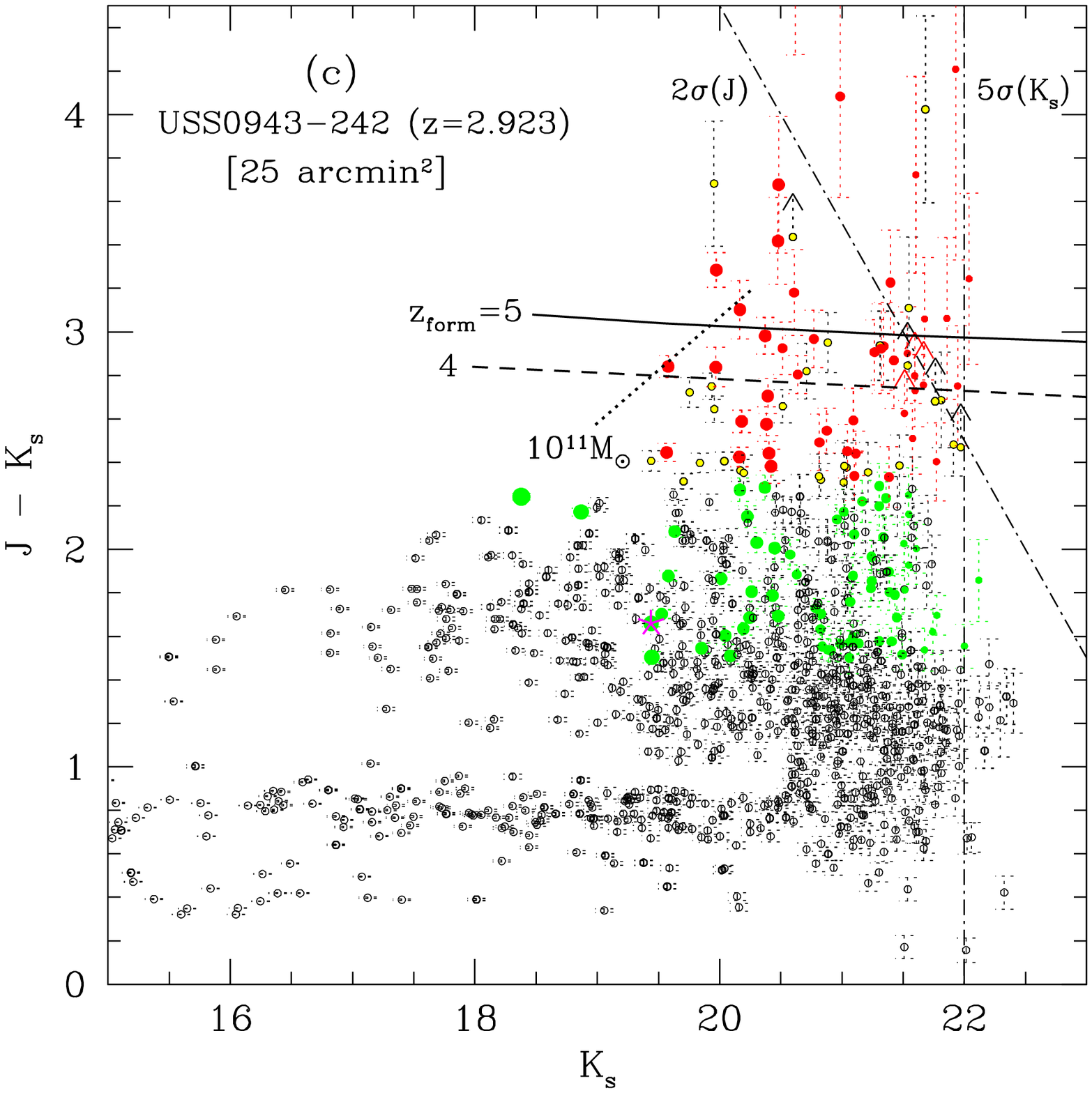}
  \epsfxsize 0.47\hsize
  \epsfbox{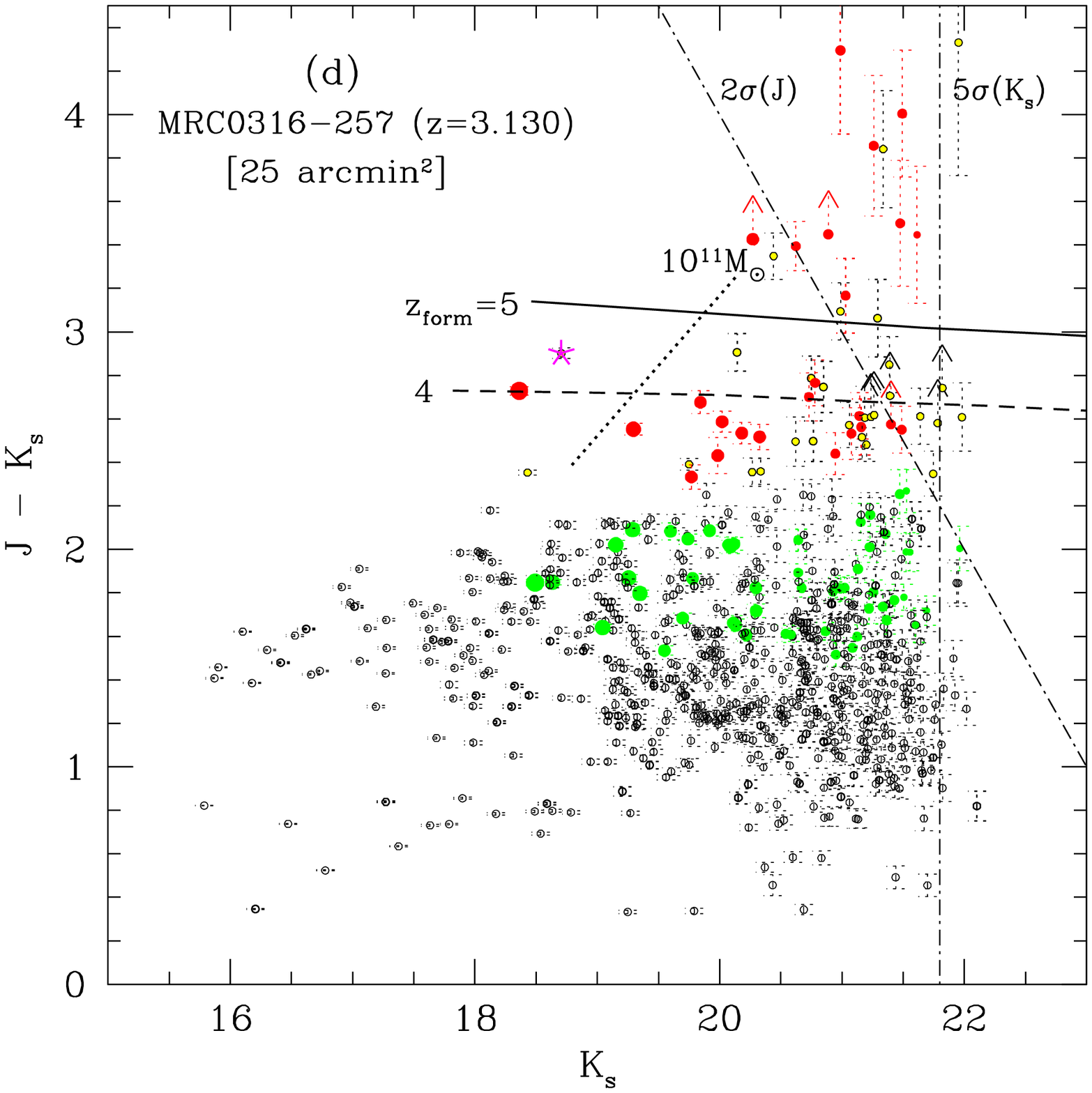}
\end{center}
\caption{
Colour--magnitude diagrams of the four proto-clusters at $2\lsim z\lsim3$.
Filled circles indicate proto-cluster member candidates selected with
$J-K_s>2.3$
(PKS~1138--262, USS~1558--003)
and $JHK_s$ diagram
(USS~0943--242, MRC~0316--257),
respectively
(see Fig.~\ref{fig:jhk} and
Kajisawa et al.\ (2006)
for our new JHK selection technique
of passive/active galaxies at $2\lsim z\lsim3$).
Colour-coding is the same as in Fig.~\ref{fig:jhk} except for the upper plots where
the red symbols indicate DRGs.
The size of the symbols is scaled according to apparent magnitudes in $K_s$-band
just for consistency with Figs.~\ref{fig:jhk} and \ref{fig:xy}.
Large star marks show the targeted radio galaxies.
The radio galaxy 1138 is extraordinary bright as it is dominated
by H$\alpha$ in $K_s$-band (Nesvadba et al.\ 2006).
Dotted error-bars show 1$\sigma$ photometric errors.
Solid, dashed and dotted lines show the expected location of colour--magnitude
relations at the radio galaxies' redshifts in the case of passive evolution \citep{kod98}. 
The iso-stellar-mass lines of 10$^{11}$M$_{\odot}$ are also shown by thick dotted lines
(the ones for 10$^{10}$M$_{\odot}$ are located at 2.5 magnitudes fainter
than those although they are not shown to keep good visibility of the plots).
The dot-dashed lines indicate 5$\sigma$ ($K_s$) and 2$\sigma$ ($J$) detection limits.
}
\label{fig:jk}
\end{figure*}

We now focus on the build-up of the red sequence in the
proto-clusters at its very first stage, which is the main theme of this paper.
Massive elliptical galaxies are seen to dominate cluster cores up to a redshift
of $z\sim1$ with the red sequence firmly in place by that epoch
(\citealp{kod98};\citealp{tan05}).
What remains uncertain is whether this holds true at higher redshift and it is
therefore unclear
at what epoch massive, cluster galaxies assembled and
from which stage they underwent mostly passive evolution.

The colour-magnitude diagram is one of the most powerful tools to investigate
galaxy formation and evolution on a large magnitude limited sample.
Fig.~\ref{fig:jk} shows NIR colour-magnitude diagrams of the four
proto-cluster fields.  Filled red circles indicate DRGs for 1138 and 1558,
and r-JHKs for 0943 and 0316, respectively.  Filled green circles are b-JHKs.
The size of these symbols correspond1s to magnitudes in $K_s$-band.
DRGs that do not satisfy the JHK criterion are shown by small filled yellow circles.
For reference, we also show the expected locations of the colour-magnitude relation
(CMR) as observed at the redshift of the central radio galaxies in the case of passive
evolution for various formation redshifts as shown.
These models are calibrated so as to reproduce the CMR of elliptical galaxies in the Coma
cluster at $z\sim0$ as a sequence of decreasing metallicity with magnitude \citep{kod98}.
A constant stellar mass of $10^{11}$\msun with Kennicut (1983) initial mass function is
approximately shown by the dotted line in each panel
\citep{kod98}.

In all the proto-cluster fields, there are a number of very red galaxies that 
have colours consistent with those of passively evolved galaxies
with high formation redshifts ($\gsim3.5$) (Kodama et al. 1998).
Such red galaxies are very rare in the general field of {\it GOODS-S} (Fig.~\ref{fig:goodss}).
Considering the magnitudes of these red galaxies
along the CMR, it becomes clear that the bright end of the red sequence
(M$_{\rm stars}>10^{11}$\msun) is already well populated by $z\sim2$ but much less
so at $z\sim3$ (Fig.~\ref{fig:jk}).
In fact, there are 12 and 4 galaxies at $z=2$ and 2.5, respectively, but there are
none or only two galaxies at $z\sim3$ above this mass limit.
Although the statistics are by all means poor at this stage, these results suggest
that the bright-end of the colour--magnitude sequence first appeared between $z=3$ and 2.
To derive any general conclusion, however, observation of a larger sample of proto-clusters
is needed so that we can average over individual characteristics of proto-clusters
even at the same epoch.

Which galaxies in the $z\sim3$ proto-clusters could be the progenitors of
the massive red sequence galaxies seen in the $z\sim2$ proto-clusters?
Where are they on the colour-magnitude diagrams at $z\sim3$?
And how is the bright end of the CMR built up between $z=3$ and 2, while only one Gyr
is available between these two epochs?
These are crucial questions to be answered to advance our understanding of
the formation of massive elliptical galaxies.
Interestingly, at $z\sim3$, there are few blue galaxies that are already
massive enough ($>10^{11}$\msun) to turn into massive red galaxies and
fill the bright-end of the CMR at $z\sim2$ simply by stopping their on-going
star formation (Fig.~\ref{fig:jk}).
Therefore, a significant growth in stellar mass is required between $z=3$ and 2
either by vigorous star formation or mass assembly by mergers or both.
In any case, to make a $10^{11}$\msun\ galaxy from a $10^{10}$\msun\ within a Gyr,
a mass growth rate of $\sim$100\msun\ per yr is required.
This would not be impossible, however.
In fact, such a high rate of star formation is often seen in submm galaxies or
dusty star-burst galaxies (such as ultra-luminous infrared galaxies and a subsample
of DRGs) found in the $2<z<3$ Universe (e.g., Chapmann et al. 2005; Webb et al. 2006).
It is also interesting to note that the peak of the submm phase and
the peak of the cosmic star formation rate just coincide with this era
(e.g., \citealp{cha05}; \citealp{bou05}).
Also, there is a dramatic evolution in dusty star formation activity in the central
radio galaxies themselves; their detection rate at submm wavelength rises from 15 per cent
at $z<2.5$ to $>75$ per cent at $z>2.5$ (Archibald et al.\ 2001; Reuland et al.\ 2004).
Therefore, it is likely that the massive galaxies are still in the
primitive phase of formation at $z\sim3$, and they are just being built up by vigorous
star formation and assembly between $z=3$ and 2.
At $z\sim2$ they eventually show up at the bright-end of the CMR at their
near-final stage of formation.
It should be noted that the fact that galaxies are on the red sequence does not
necessarily mean that star formation is completely switched off in these systems 
(Papovich et al.\ 2006; Kriek et al.\ 2006).
In fact, a large fraction of DRGs at $z>2$ tend to contain some on-going star formation
at the level of a few to a few tens of solar masses per year, but they are reddened
by dust extinction (Webb et al. 2006).  We note however that those DRGs are found in
random general fields such as {\it GOODS} rather than known proto-clusters as we
targeted in this study, and it is not {\it a priori} clear
if such remaining star formation is also present in the red massive galaxies in our
proto-clusters or not.

Some of the r-JHKs and b-JHKs are bright enough to be targeted in spectroscopy
both in the optical and in the NIR in order to confirm
their physical association to the proto-clusters around the radio galaxies.
This is the most important next step as it will also confirm the interesting results 
presented here on the first appearance of massive red sequence galaxies
in the proto-clusters at $2<z<3$.

%
%
\section{Spatial distribution of the proto-cluster members}
\label{sec:xy}

\begin{figure*}
\begin{center}
\leavevmode
  \epsfxsize 0.47\hsize
  \epsfbox{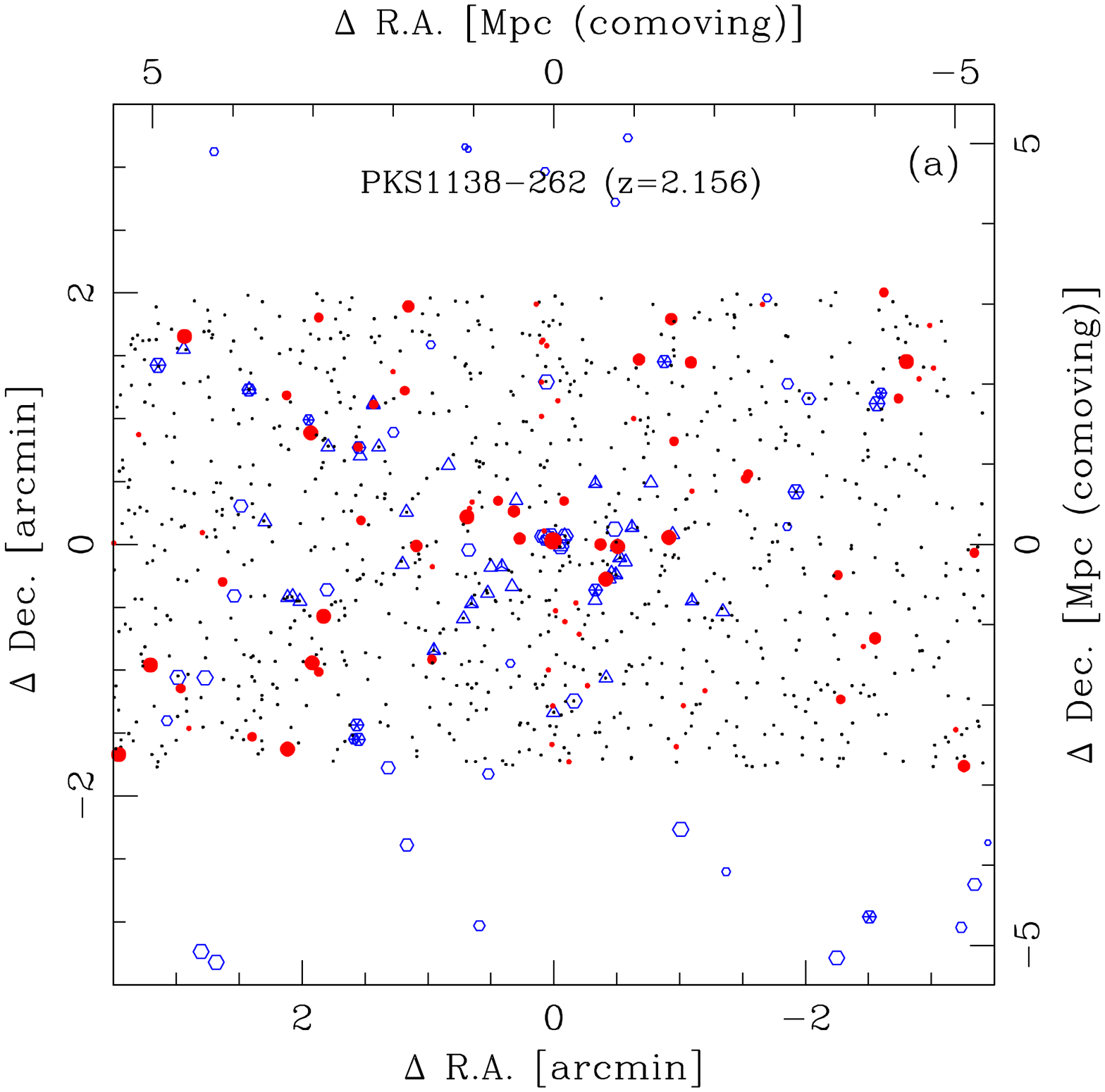}
  \epsfxsize 0.47\hsize
  \epsfbox{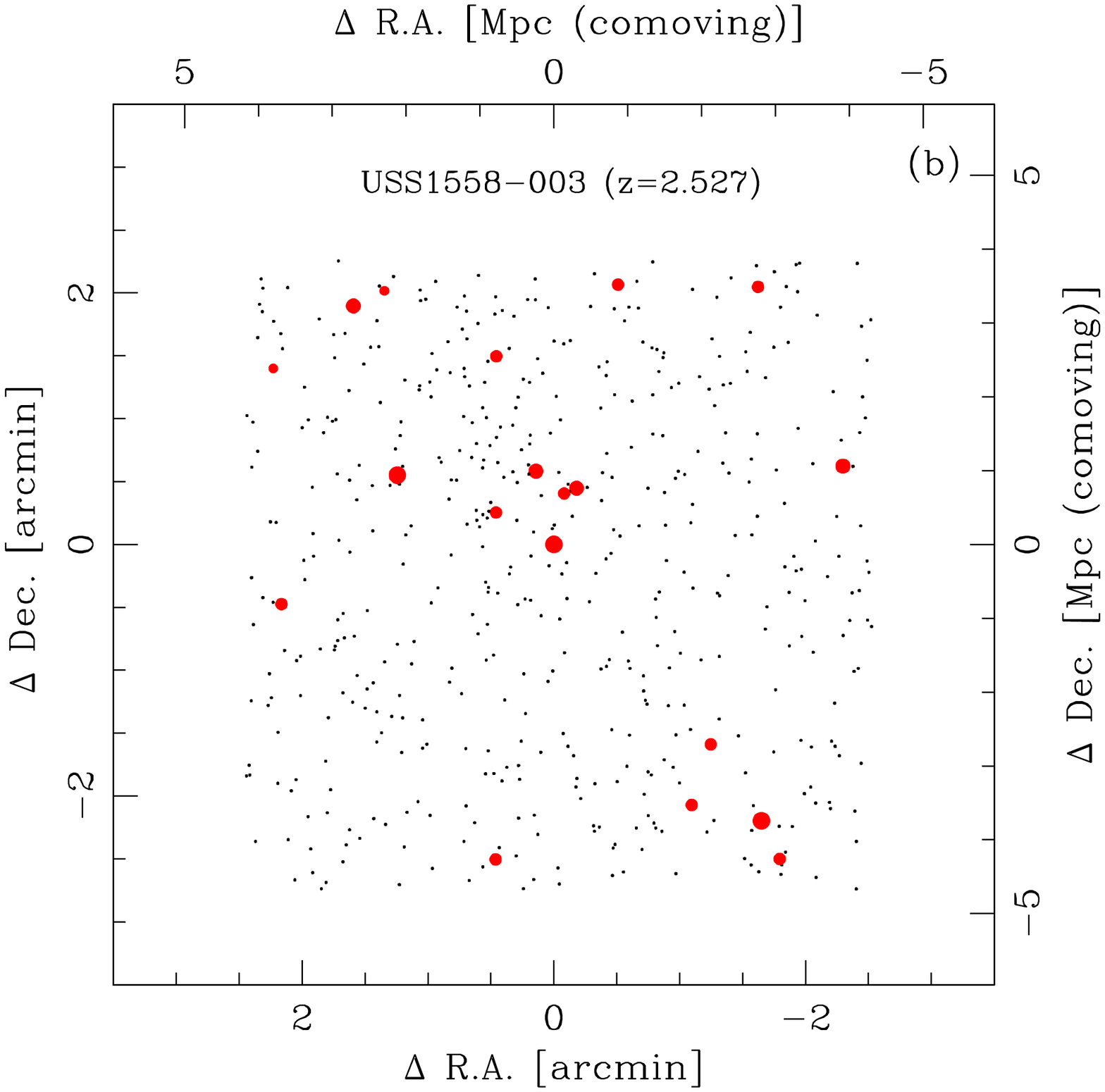}\\
  \hspace*{-0.1cm}
  \epsfxsize 0.47\hsize
  \epsfbox{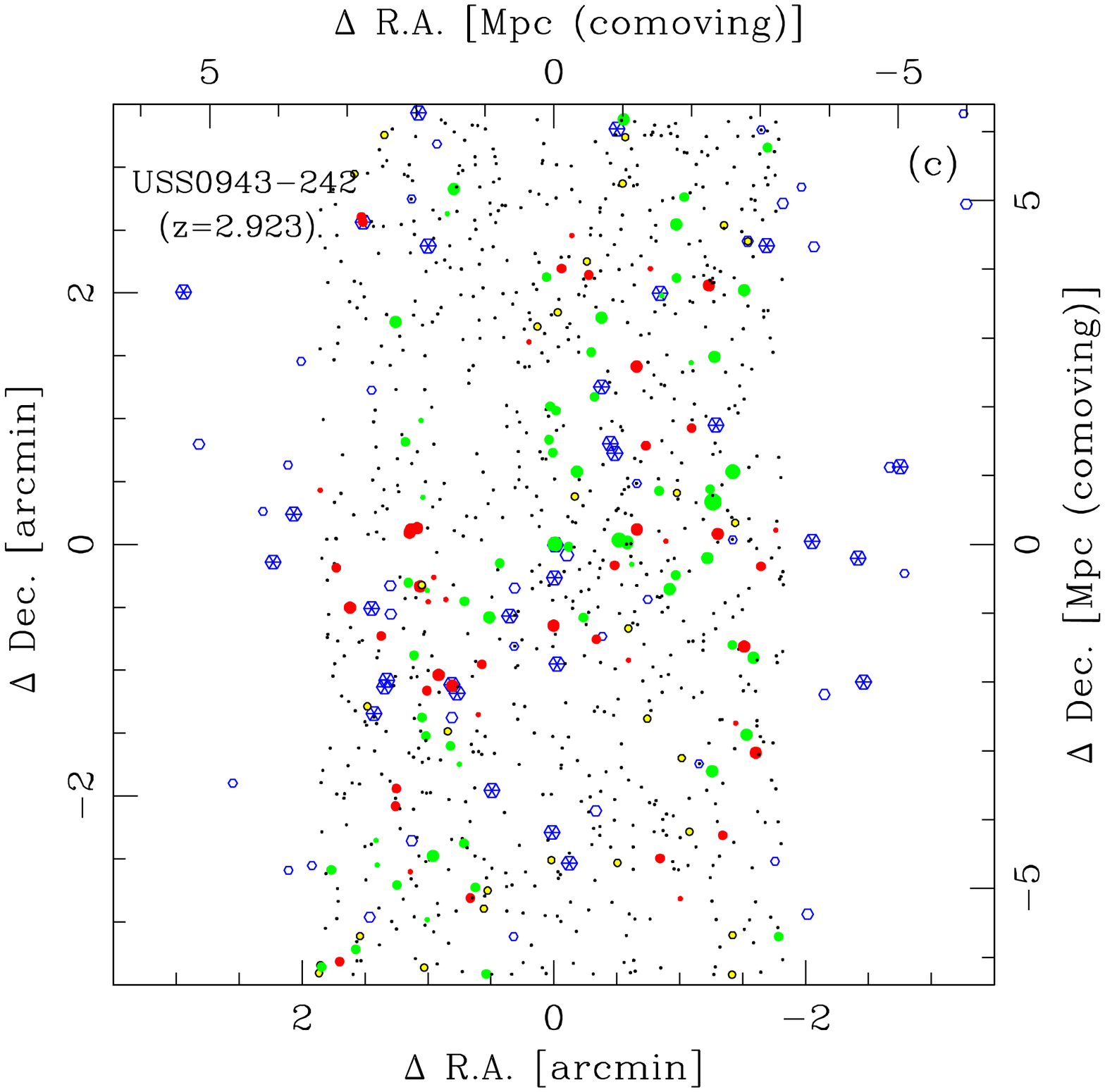}
  \epsfxsize 0.47\hsize
  \epsfbox{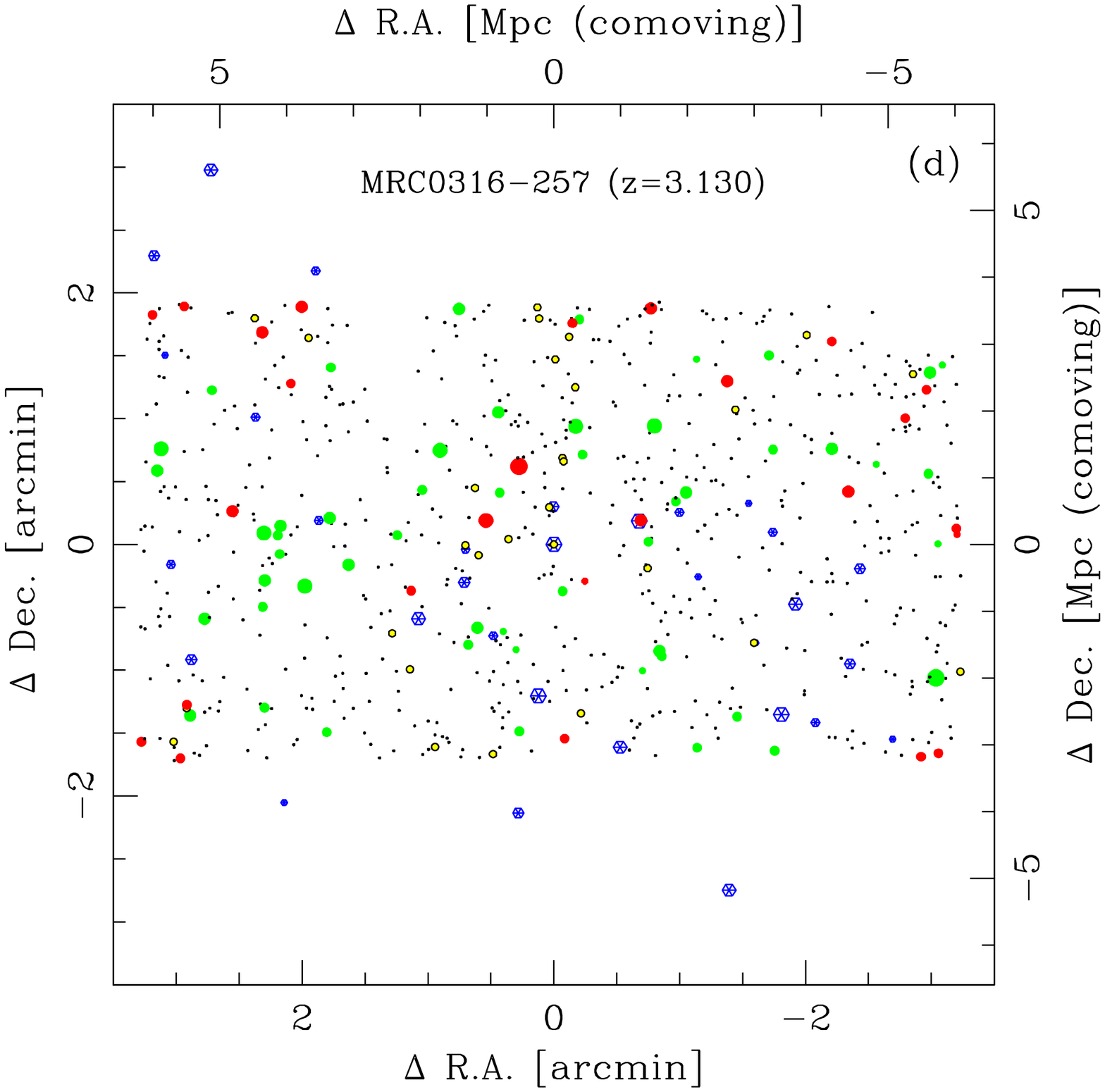}
\end{center}
\caption{
Spatial distribution of the galaxies in the
four
proto-clusters at $2<z<3$.
The radio galaxies are located at the centre of the fields and the coordinates
are given with respect to the radio galaxies in arcmin and Mpc (co-moving).
Filled circles indicate proto-cluster member candidates selected with
$J-K_s>2.3$ (PKS~1138--262, USS~1558--003) and $JHK_s$ diagram
(USS~0943--242, MRC~0316--257), respectively.
Colour-coding is the same as in Fig.~\ref{fig:jhk} except for the upper plots where
the red symbols indicate DRGs.
The size of these symbols is scaled according to apparent magnitudes in $K_s$-band.
Open hexagons and open triangles are Ly$\alpha$ and H$\alpha$ emitters, respectively,
and those with a star mark are spectroscopically confirmed emitters associated to the
radio galaxies.
The size of the latter symbols is scaled according to line flux.
}
\label{fig:xy}
\end{figure*}

The spatial structure of proto-clusters is another interesting
subject, since it reflects the primordial density fluctuation and the
early stage of structure formation.  \cite{kaj06} have shown a hint of
filamentary structure in the NIR-selected galaxies in two
proto-clusters at $z\sim2.5$, even within the $1.6\times1.6$
arcmin$^2$ field of view.  With the larger FoVs of MOIRCS and SOFI, we
are able to trace the proto-clusters structures to much larger
extents, which will demonstrate any filaments if present with more
certainty.

The distribution of galaxies in our four proto-clusters is shown in
Fig.~\ref{fig:xy}. Filled circles indicate the NIR selected galaxies,
while open symbols show the \lya\ and/or \ha\ emitters. See the
caption for details.  The spatial distribution of the NIR selected
galaxies is not uniform in general.  For example, brighter DRGs in
1138 seem to form an elongated filament in the East-West direction,
especially close to the radio galaxy.
\cite{cro05} also found a similar elongated structure for X-ray
emitting galaxies in the field of 1138.
The JHKs and DRGs in 0943 are
another good example, as they show a large overall elongated structure
from ESE to NW.  The filaments traced by the NIR-selected galaxies are
roughly in the same direction as those formed by the \lya\ and/or \ha\
galaxies.
These non-uniform structures
strongly indicate that these proto-clusters
are right at the stage of initial vigorous assembly to form the basic
shapes of the clusters.

To be more quantitative, we have performed a 2D Kolmogorov-Smirnoff (K-S) test, and found
that the probabilities that the DRGs ($K_s<20.5$) in 1138 and JHKs ($K_s<21$) in
0943 are drawn from a random distribution are only 0.05 and 0.0006, respectively.
In contrast, probabilities that the NIR-selected galaxies and the emitters have
the same spatial distribution are 0.24 and 0.09 for 1138 and 0943, respectively.
This suggests that the DRGs and JHKs are clustered and possibly in a similar fashion
as the line emitters at the redshifts of central radio galaxies.
Note that we have subtracted field contamination for the NIR-selected sample
using the {\it GOODS-S}
\footnote{As a normal procedure of the 2D K-S test, for each galaxy we divide the
survey area into four quadrants centred on the galaxy and count the number of galaxies
in each quadrant.  Here we make a field subtraction by calculating the expected
number of field galaxies in each quadrant using a simple scaling of the field sample
by its area.}.

Although the NIR-selected galaxies and the \lya/\ha\ emitters trace
similar structures, it is noteworthy that the individual galaxies
rarely overlap between these two populations.  This indicates that
most of the emitters are not massive enough to be detected in our NIR
imaging ($M_{\rm stars}\lsim 10^{10}$\msun), and at the same time,
even the b-JHK (with some on-going star formation) in the $z\sim3$
proto-clusters do not show strong enough Ly$\alpha$ emission to be
detected in the narrow-band emitter surveys
($L$ (Ly$\alpha$) $\lsim 6.2 \times 10^{41}$ erg s$^{-1}$ for 0943 and
$6.9 \times 10^{41}$ erg s$^{-1}$ for 0316 at 5 sigma).
There is also a
possibility that the redshifts of some of these b-JHKs may be offset
by more than $\sim$ 1500 km s$^{-1}$ so that they cannot be detected
in the narrow bands.  In any case, these proto-clusters consist of
both massive evolved populations as traced by JHKs or DRGs, and less
massive, young populations as traced by \lya/\ha\ emitters.

%
%
\section{Summary}
\label{sec:summary}

We targeted proto-clusters around four radio galaxies at $2\lsim z\lsim3$
using wide-field NIR instruments.
Most of these fields are known to show a large number of Ly$\alpha$
and/or H$\alpha$ emitters at the same redshifts as the radio galaxies.
We see a clear excess of NIR selected galaxies (JHK-selected galaxies
as well as DRG) in these fields, which confirms that they indeed
contain proto-clusters with not only young emitters but also evolved
galaxy populations.  The spatial distribution of these NIR selected
galaxies is filamentary, similar to the structures traced by the
emitters.  The observed difference in the number of red sequence
galaxies between the $z \sim 3$ and the $z \sim 2$ fields in our
sample may imply that
the bright end of the red sequence first appeared between $z=3$ and 2.

Since we are targeting the biased high density regions,
one could imagine that galaxy formation and evolution 
take place faster here than in lower density or randomly selected
regions (e.g., Cen \& Ostriker 1993; Diaferio et al.\ 2001;
Tanaka et al.\ 2005; Steidel et al.\ 2005).
Stellar mass functions derived for `general' fields such as
{\it GOODS-MUSIC} (Fontana et al. 2006) and {\it FDF+GOODS-S} (Drory et al. 2005)
show that the massive end ($>$10$^{11}$\msun) starts to deviate
from the low-$z$ ($z\sim0.5$) function significantly at $z\sim1.5$ or so,
which is slightly later than the epoch when we see the truncation of
the massive-end of cluster red sequence ($z\sim2.5$).
This could be suggestive that massive galaxies in proto-clusters are
assembled a bit earlier than those in the general fields.
However, a quantitative statement on environmental dependence will have to wait
until stellar mass functions for a range of redshift can be constructed from
observations of a larger sample of proto-clusters, which can be compared directly
to the stellar mass functions of field galaxies.

%
%
\section*{Acknowledgements}
We thank the anonymous referee for careful reading of the draft and making
some useful comments, which improved the paper.
This work was financially supported in part by a Grant-in-Aid for the
Scientific Research (No.\, 18684004) by the Japanese Ministry of Education,
Culture, Sports and Science.
TK acknowledges hospitality of ESO during his stay in 2005 when a part of this
work was done. JK is supported by SFB 439 of the DFG, the German Science Foundation.
This study is based on data collected at Subaru Telescope, which is operated by
the National Astronomical Observatory of Japan.
This study is also in part based on data collected at New Technology
Telescope at the ESO La Silla Observatory under Program ID: 076.A-0670
(for USS~1558--003) and the Very Large Telescope at the
ESO Paranal Observatory under Program ID: LP168.A-0485 (for GOODS-S).

%
%

\end{document}